\documentclass[twocolumn]{emulateapj}
\usepackage{amsmath}
\usepackage{lineno}

\newcommand{\nstars}{2476}
\newcommand{\nsmallstars}{1630}

\begin{document}
\bibliographystyle{apj}

\title{Giant planet occurrence within 0.2 AU of low-luminosity red giant branch stars with K2}

\author{Samuel K.\ Grunblatt\altaffilmark{1,*}}

\shorttitle{Close-in Giant Planets Around Red Giant Stars}
\shortauthors{Grunblatt et al.}

\author{Daniel Huber\altaffilmark{1}}
\author{Eric Gaidos\altaffilmark{2}}
\author{Marc Hon\altaffilmark{3}}
\author{Joel C. Zinn\altaffilmark{4}}
\author{Dennis Stello\altaffilmark{3,5,6}}

\altaffiltext{1}{Institute for Astronomy, University of Hawaii,
2680 Woodlawn Drive, Honolulu, HI 96822, USA}

\altaffiltext{2}{Department of Earth Sciences, University of
Hawaii at Manoa, Honolulu, Hawaii 96822, USA} 
\altaffiltext{3}{School of Physics, University of New South Wales, NSW 2052, Australia}
\altaffiltext{4}{Department of Astronomy, The Ohio State University, Columbus, OH 43210}
\altaffiltext{5}{Sydney Institute for Astronomy, School of Physics, University of Sydney, NSW 2006, Australia}
\altaffiltext{6}{Stellar Astrophysics Centre, Department of Physics and Astronomy, Aarhus University, Ny Munkegade 120, DK-8000 Aarhus C, Denmark}

\altaffiltext{*}{skg@ifa.hawaii.edu}

\begin{abstract}

Every Sun-like star will eventually evolve into a red giant, a transition which can profoundly affect the evolution of a surrounding planetary system. The timescale of dynamical planet evolution and orbital decay has important implications for planetary habitability, as well as post-main sequence star and planet interaction, evolution and internal structure. Here, we investigate these effects by estimating planet occurrence around \nstars\ low-luminosity red giant branch (LLRGB) stars observed by the NASA \emph{K2} mission. We measure stellar masses and radii using asteroseismology, with median random uncertainties of 3.7\% in mass and 2.2\% in radius. We compare this planet population to the known population of planets around dwarf Sun-like stars, accounting for detection efficiency differences between the stellar populations. We find that 0.51\% $\pm$ 0.29\% of LLRGB stars host planets larger than Jupiter with orbital periods less than 10 days, tentatively higher than main sequence stars hosting similar planets (0.15\% $\pm$ 0.06\%). Our results suggest that the effects of stellar evolution on the occurrence of close-in planets larger than Jupiter is not significant until stars have begun ascending substantially up the red giant branch ($\gtrsim$ 5-6 R$_\odot$).




\end{abstract}

\section{Introduction}

As a star like our Sun ages, changes in stellar luminosity, composition and structure can induce changes in orbiting planets \citep{villaver2014, veras2016}. The increase in stellar irradiation during the red giant phase of stellar evolution may lead to planet inflation \citep{guillot1996, lopez2016}. Tides in both the star and the planet can also affect planet interiors, causing inflation and disruption of their magnetic dynamo \citep{bodenheimer2001, driscoll2015}. However, despite being relatively luminous, and thus overrepresented in magnitude-limited surveys \citep{malmquist1922}, the variability of evolved stars makes it difficult to detect transiting planets around them \citep{sliski2014}. Therefore even though these systems hold many insights into the nature of planet inflation, migration and evolution, transiting planet surveys have largely avoided these stars.

Previous searches for planets around evolved stars utilized radial velocity measurements  \citep{hatzes2000,sato2005,reffert2015}. Despite the relatively long history of planet searches around evolved stars, no planets were found interior to 0.5 AU around such stars, suggestive of intrinsic differences between the main sequence and evolved system populations \citep{johnson2010, jones2016}. The recent explosion in planet discoveries around Sun-like and smaller stars fueled by transit surveys has been accompanied by only a handful of planet transit detections around evolved stars \citep{lillo-box2014, barclay2015, vaneylen2016, grunblatt2016, grunblatt2017}. To determine whether the relatively small number of planets known around evolved stars is due to small survey size, planet detection difficulties unique to evolved stars, or an intrinsic lack of planets, a systematic transit survey of evolved stars is needed.

Here, we investigate over 10 000 stars observed by the \emph{K2} mission \citep{howell2014} to estimate planet occurrence around low-luminosity red giant branch stars. Searching for planet transits around these moderately evolved stars captures the intrinsic photometric variability due to the oscillations of these stars as well. These oscillations can be used to measure stellar densities and surface gravities through asteroseismology, which we use to calculate planet occurrence statistics with more precision than current spectroscopic techniques would allow \citep{huber2013,petigura2017}. We restrict our sample to \nstars\ of these stars whose radii are large enough for precise characterization with asteroseismology but are also small enough to allow planet transit detection using the 30-minute cadence data of \emph{K2}. We use this sample to determine planet occurrence for our evolved stars, which we compare to planet occurrence estimates around main sequence stars.


\section{Target Selection}

\begin{figure}[t!]
\epsscale{1.3}
\plotone{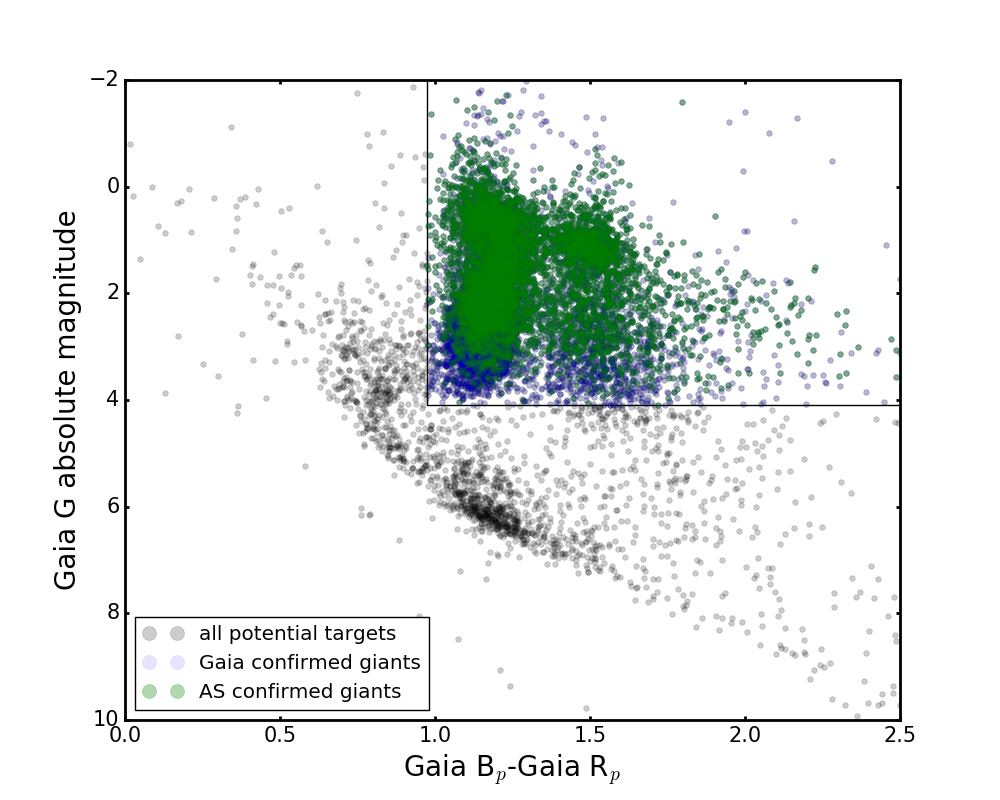}
\caption{Color-magnitude diagram made using \emph{Gaia} Data Release 2 data. We restrict our asteroseismic analysis to those stars with colors and magnitudes consistent with giant stars (colored points, above and right of black lines). Particularly red targets seen here are due to high reddening fields in \emph{K2}. Targets detected as oscillating giants by multiple asteroseismic pipelines are shown in green. \label{gaiacolors}}
\end{figure}

\begin{figure*}[ht!]
\epsscale{1.15}
\plottwo{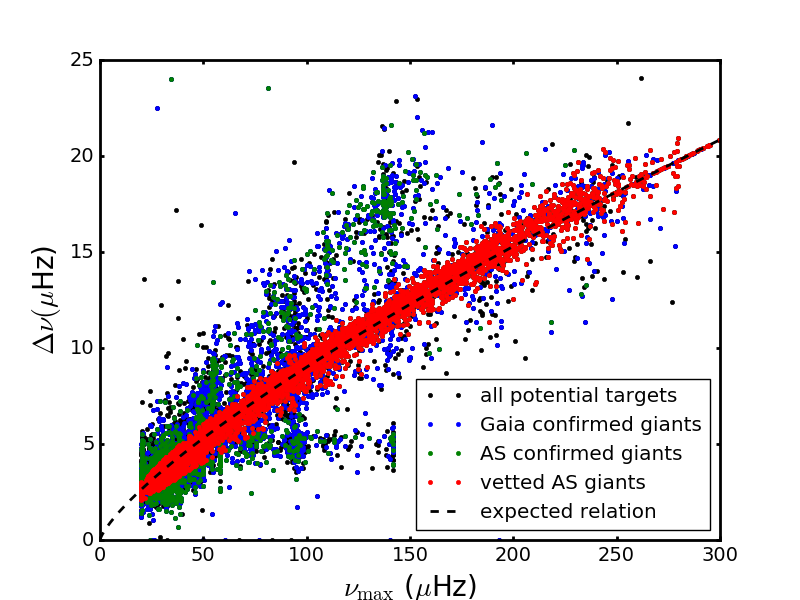}{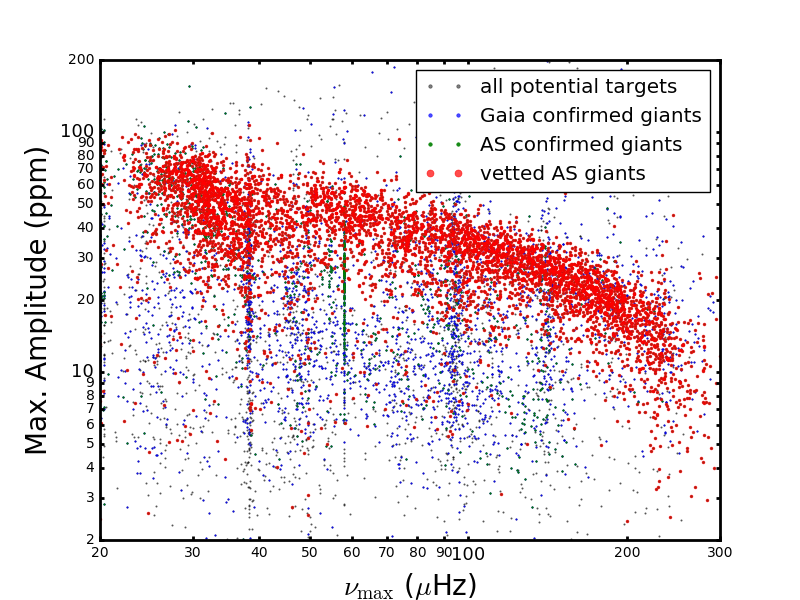}
\caption{Left: Frequency of maximum oscillation power $\nu_\mathrm{max}$ versus large frequency separation $\Delta\nu$. Asteroseismic oscillating giants passing all of our quality cuts are shown in red, while rejected stars are plotted in blue, black and green. The dotted line corresponds to the previously published power-law relation between $\nu_{\mathrm{max}}$ and $\Delta\nu$ \citep{stello2009,yu2018}. Right: Maximum oscillation amplitude in ppm versus frequency of maximum oscillation power $\nu_\mathrm{max}$. Helium-burning stars in the red clump can be seen as an increase in amplitude dispersion between 20 and 40 $\mu$Hz.
\label{numaxdnu}}
\end{figure*}

The targets for our study were chosen as follows:
 
\begin{enumerate}
    \item 10 444 initial targets observed for this study were selected by the Giants Orbiting Giants \emph{K2} Guest Observer campaigns (GO4089, GO5089, GO6084, GO7084, GO8036, GO10036, GO11048, GO12048, GO13048, GO14004, GO15004, GO16004, PI: D. Huber). These stars were identified as having temperatures between 4500 and 5500 K, surface gravities of 2.9 $<$ log $g$ $<$ 3.5, and magnitudes of K$_p$ $<$ 15 as compiled in the Ecliptic Plane Input Catalog \citep[EPIC;][]{hipparcos1997,gaia2018,apogee2017,rave2017,lamost2012,huber2016} to increase the likelihood that stellar oscillations would be detectable by \emph{K2} \citep{chaplin2014, stello2015}. 
    \item 458 additional stars observed serendipitously as part of the \emph{K2} Galactic Archaeology Program \citep[GAP,][]{stello2017} were identified as potential LLRGB stars using stellar radii determined using EPIC parameters \citep{huber2016}. Including these stars with EPIC radii between 3 and 10 R$_\odot$ increases our target sample to a total of 10902 stars (Figure \ref{gaiacolors}).
    \item After \emph{Gaia} Data Release 2 became available \citep{gaia2018}, stars with absolute magnitudes \emph{Gaia} {\it G} magnitude $>$ 4.1 and \emph{Gaia} {\it B}$_p$-{\it R}$_p$  $<$ 0.9 and $>$ 3.0 were excluded from our study. These cuts were empirically determined by the authors through a visual analysis of Figure \ref{gaiacolors} to isolate red giant branch stars and remove main sequence dwarfs and subgiants. Spot checks of the light curves of stars outside this region revealed no detectable asteroseismic signals. Thus this left 8933 potential oscillating red giant stars (blue and green points, upper right of Figure \ref{gaiacolors}), a rejection of 18\% of stars in our sample. As the original \emph{Kepler} mission had a smaller contaminating fraction of giants but a larger contaminating fraction of subgiants as determined from \emph{Gaia} photometry \citep{berger2018}, and its targets were chosen using similar photometric parameter determination, 18\% is within the range of expected contamination rates.
    \item Multiple asteroseismic pipelines were then used to ensure that oscillations could be detected unambiguously and could be used to accurately characterize the host star \citep[]{huber2009, hon2018, zinn2019}, leaving 6330 oscillating stars in the sample (green points in Figures \ref{gaiacolors} and \ref{numaxdnu}). We then performed additional vetting based on the quality of the observed oscillations and stellar parameters determined therewith (see Figure \ref{numaxdnu}, Section 3.1).
    
\end{enumerate}

\section{Asteroseismology}

\subsection{Data Analysis}

Asteroseismology is the study of relating observed oscillations to the physical properties of a star \citep{christensen1983}. These oscillations can be seen in the power spectra of stellar light curves. Numerous analysis packages have been developed to derive stellar properties accurately and precisely from asteroseismic oscillation signals, by analysis of power spectra of oscillating stars \citep{huber09, hon2018}. In order to accurately and precisely determine the stellar radii and masses of all the stars in our sample, we produce power density spectra of all of our targets from their \emph{K2} lightcurves created with the K2SFF algorithm \citep{vanderburg2014}. 

Known \emph{K2} lightcurve features, such as those produced by the firing of thrusters to keep the spacecraft pointing accurate every 6 hours, can mimic an asteroseismic signal. In addition, astrophysical false positive signals can also be produced by eclipsing binary systems or classical pulsators such as RR Lyrae variables. To exclude these unwanted signals from our analysis, we median filter our light curves with a 3-day window in addition to the initial detrending done by the K2SFF algorithm. We also exclude data within 1 day of any gap in data acquisition within a campaign, as well as within 1 day of the start and end of each campaign to remove spurious signals near stellar oscillation or transit timescales. 

In order to determine whether oscillations were present in a particular stellar light curve, we use deep learning-based classification to detect oscillations from 2-dimensional images of power spectral density plots of \emph{K2} light curves following the method of \citet{hon2018}. This technique is trained using \emph{Kepler} data curated by asteroseismic experts to assign a probability $p$ that a star is or is not oscillating. Though \citet{hon2018} achieved an accuracy over 98\% on their test sample using a threshold probability of $p_\mathrm{thres}$ $>$ 0.58, we adopt a more conservative threshold of $p_\mathrm{thres}$ $>$ 0.95 in our final analysis to ensure minimal contamination by false positives in our dataset. We also apply the Bayesian classification scheme of \citep{zinn2019} to our light curve data to classify the star as oscillating or not. We find that our classification of asteroseismically oscillating stars agree between these asteroseismic pipelines in more than 99\% of cases.

We then perform an asteroseismic analysis on all power spectra that pass the filters above, calculating the best-fit frequency of maximum power ($\nu_\mathrm{max}$) and regular frequency spacing ($\Delta\nu$) between sequential radial oscillation modes using the \citet{huber2009} pipeline, which has been well established for the asteroseismic analysis of \emph{Kepler} and \emph{K2} photometry \citep{huber2011, huber2013,stello2017}. We calculate uncertainties for our asteroseismic quantities using a Monte Carlo method, producing 100 realizations of each asteroseismic fit and using the standard deviation of the sample of asteroseismic fits for each star to determine parameter uncertainties as described in \citet{huber2011}. We then use these $\nu_\mathrm{max}$ and $\Delta\nu$ uncertainties to determine uncertainties on stellar mass and radius. We cross check our $\nu_\mathrm{max}$ results with two other asteroseismic pipelines \citep{hon2018, zinn2019} and find that our $\nu_\mathrm{max}$ estimates agree within 1\% on average, and more than 95\% of stars designated as oscillating have $\nu_\mathrm{max}$ values that agree to within 5\%. We reject all stars which do not meet these requirements, resulting in a sample of 6330 oscillating stars (green points in Figures \ref{gaiacolors} and \ref{numaxdnu}).

We remove additional poor asteroseismic detections by excluding stars which have a measured $\nu_{\mathrm{max}}$ above 285 $\mu$Hz, below 20$\mu$Hz, or within 0.05 $\mu$Hz of 58.05 $\mu$Hz due to an observed nonphysical pileup of $\nu_{\mathrm{max}}$ values observed at this frequency (see \citet{zinn2019} for a more detailed analysis of this feature). Visual inspection of stars showing $\nu_{\mathrm{max}}$ values within this range reveal stellar power spectra polluted by a periodic signal not linked to stellar oscillation, and thus these stars are excluded from our subsequent analysis. We note the low frequency harmonic of this feature at $\approx$ 39 $\mu$Hz, but do not mask this feature as stars with similar $\nu_{\mathrm{max}}$ values reside in the red clump or are too large to allow planet transit detection. In addition, we reject stars whose $\nu_{\mathrm{max}}$ and $\Delta\nu$ values disagree with the empirical relation derived by \citet{yu2018} (given in the following paragraph) by more than 20\%. This leaves us with a vetted asteroseismic sample of 5227 oscillating stars (red points in Figure \ref{numaxdnu}).

The left panel of Figure \ref{numaxdnu} illustrates the relation between $\nu_{\mathrm{max}}$ and $\Delta\nu$ for stars in our target sample including both the stars which pass our asteroseismic vetting (red) as well as those which do not (green), including those designated as dwarfs by \emph{Gaia} photometry (black) and those without consistent oscillations found by multiple asteroseismic pipelines (blue). The right panel gives the correlation between maximum oscillation amplitude and $\nu_{\mathrm{max}}$ for all stars in our target sample. We highlight the \citet{yu2018} relation determined between $\nu_{\mathrm{max}}$ and $\Delta\nu$ from a sample of 16094 \emph{Kepler} red giants,

\begin{equation}
\Delta\nu = \alpha (\nu_\mathrm{max})^\beta,
\end{equation}

where $\alpha = 0.267$ and $\beta = 0.764$. We also note the pile up of measured $\nu_{\mathrm{max}}$ values at the known \emph{K2} thruster firing frequency of 47 $\mu$Hz and its multiples. However, stars with measured $\nu_{\mathrm{max}}$ values near these thruster harmonics do not seem to be preferred by our three tested pipelines, and thus we do not mask these stars from our analysis. We also note that $\nu_{\mathrm{max}}$ values near 283 $\mu$Hz calculated by our pipelines tend to be inaccurate due to the reflection of both sub- and super-Nyquist oscillation peaks about the Nyquist frequency, causing an artificial oscillation peak at the Nyquist frequency for stars oscillating slightly above or below this value \citep{yu2016}. However, since $\Delta\nu$ is still well-constrained for these stars, we use the derived relation of \citet{yu2018} to estimate $\nu_\mathrm{max}$ analytically for all stars with $\nu_\mathrm{max}$ $>$ 280 $\mu$Hz, which we then use to derive stellar masses and radii. We note that this relation assumes a fixed mass for these stars, but as we are investigating planet occurrence as a function of stellar radius and not stellar mass in this sample, the inaccuracy of these stellar masses will not influence our planet occurrence results. 





\begin{deluxetable*}{cccccc}
\tabletypesize{\scriptsize}
\tablecaption{Asteroseismic Parameters\label{table1}}
\tablewidth{0pt}
\tablehead{
\colhead{EPIC ID} & \colhead{$\nu_{\rm max}$ ($\mu$Hz)} & \colhead{$\Delta\nu$ ($\mu$Hz)} & \colhead{Stellar Radius (R$_\odot$)} & \colhead{Stellar Mass (M$_\odot$)} & \colhead{$T_\mathrm{eff}$\tablenotemark{a}}
}
\startdata

201091253 & 116.9 $\pm$ 0.4 & 10.39 $\pm$ 0.09 & 5.66 $\pm$ 0.21 & 1.11 $\pm$ 0.04 & 4916 \\
201092039 & 160.1 $\pm$ 1.7 & 13.87 $\pm$ 0.17 & 4.37 $\pm$ 0.18 & 0.90 $\pm$ 0.03 & 4946 \\
201102783 & 60.2 $\pm$ 0.7 & 6.23 $\pm$ 0.05 & 7.78 $\pm$ 0.09 & 1.07 $\pm$ 0.03 & 4794 \\
201106507 & 190.6 $\pm$ 1.1 & 15.86 $\pm$ 0.10 & 4.07 $\pm$ 0.06 & 0.95 $\pm$ 0.03 & 5377\\
201114106 & 220.3 $\pm$ 7.9 & 17.69 $\pm$ 0.15 & 3.77 $\pm$ 0.18 & 0.95 $\pm$ 0.09 & 5068\\
201134999 & 83.2 $\pm$ 1.5 & 8.25 $\pm$ 0.15 & 6.86 $\pm$ 0.39 & 1.19 $\pm$ 0.06 & 5100\\
201145260 & 132.8 $\pm$ 1.2 & 11.75 $\pm$ 0.11 & 5.07 $\pm$ 0.12 & 1.03 $\pm$ 0.03 & 5070\\
201145884 & 126.7 $\pm$ 0.8 & 10.99 $\pm$ 0.16 & 5.56 $\pm$ 0.36 & 1.17 $\pm$ 0.12 & 4961\\
201161185 & 70.8 $\pm$ 1.5 & 7.91 $\pm$ 0.11 & 6.28 $\pm$ 0.12 & 0.84 $\pm$ 0.03 & 4915\\
201195238 & 64.1 $\pm$ 0.6 & 6.61 $\pm$ 0.05 & 7.83 $\pm$ 0.12 & 1.19 $\pm$ 0.03 & 5000\\
etc.

\enddata 
\tablenotetext{a}{Uncertainties on T$_\mathrm{eff}$ are 94 K for all stars in our sample, based on the \citet{gonzalez2009} color-T$_\mathrm{eff}$ relation used in this analysis.}

\tablecomments{The full results of our asteroseismic analysis are available for download with the electronic version of this Article. We include parameters for all \nstars\ stars selected for our occurrence analysis.}

\end{deluxetable*}

\begin{figure*}[ht!]
\epsscale{1.17}
\plottwo{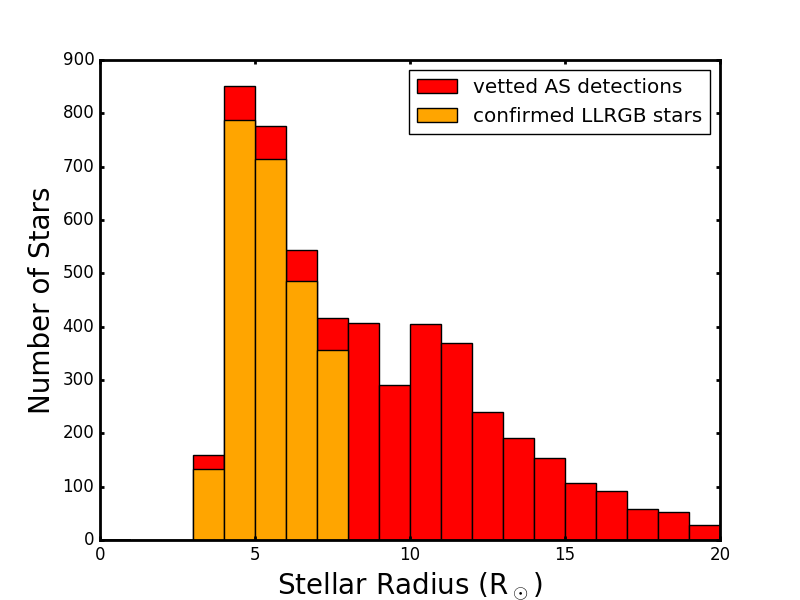}{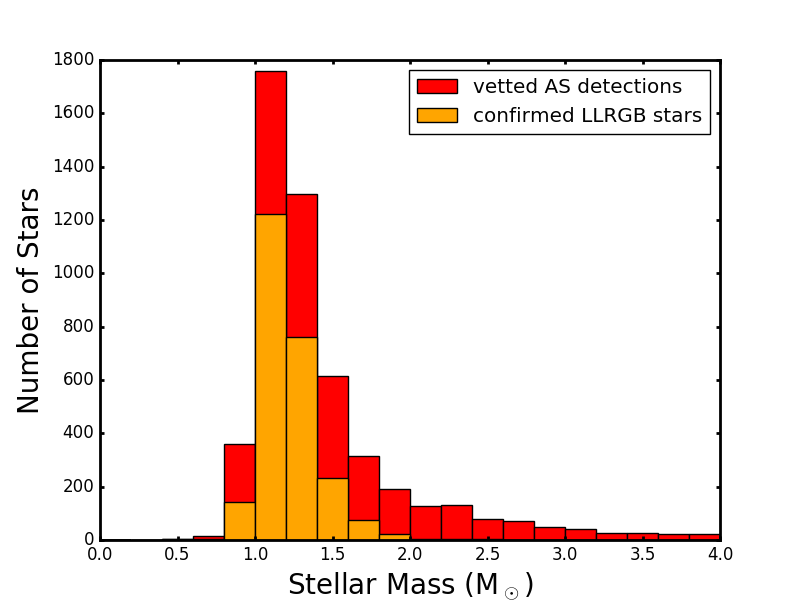}
\caption{Asteroseismic radius (left) and mass (right) distribution of our target sample. Stars which pass our asteroseismic vetting (\S 3.1) are shown in red. \nstars\ stars shown in orange have radii $<$ 8 R$_\odot$, and pass our quality cuts into our defined range of low-luminosity red giant branch (LLRGB) stars. \label{starrads}}
\end{figure*}

\subsection{Stellar Radius Determination}

To estimate stellar masses and radii from our measured $\nu_\mathrm{max}$ and $\Delta\nu$ values which passed our asteroseismic vetting (Figure \ref{numaxdnu}, red points), we use the asteroseismic scaling relations of  \citet{brown1991} and \citet{kjeldsen1995}:

\begin{equation}
\frac{\Delta \nu}{\Delta \nu_{\odot}} \approx f_{\Delta \nu} \left(\frac{\rho}{\rho_{\odot}}\right)^{0.5} \: ,
\end{equation}

\begin{equation}
\frac{\nu_{\rm max}}{\nu_{\rm max, \odot}} \approx \frac{g}{{\rm g}_{\odot}} \left(\frac{T_{\rm eff}}{T_{\rm eff, \odot}}\right)^{-0.5} \: .
\end{equation}

\noindent
where $f_{\Delta\nu}$ is the correction factor suggested by \citet{sharma16} to account for known deviations from the previously established asteroseismic scaling relation. Equations (1), (2) and (3) can be rearranged to solve for mass and radius:

\begin{equation}
\frac{M}{\rm {\rm M}_\odot}   \approx   \left(\frac{\nu_{\rm
max}}{\nu_{\rm max,
\odot}}\right)^{3}\left(\frac{\Delta \nu}{f_{\Delta \nu}
\Delta \nu_{\odot}}\right)^{-4}\left(\frac{T_{\rm eff}}{T_{\rm
eff, \odot}}\right)^{1.5}   
\end{equation}

\begin{equation}
\frac{R}{\rm R_\odot}   \approx  \left(\frac{\nu_{\rm
max}}{\nu_{\rm max, \odot}}\right)\left(\frac{\Delta
\nu}{f_{\Delta \nu} \Delta \nu_{\odot}}\right)^{-2}\left(\frac{T_{\rm
eff}}{T_{\rm eff, \odot}}\right)^{0.5}.
\end{equation}


We combine our $\nu_\mathrm{max}$ and $\Delta\nu$ values calculated via the \citet{huber2009} pipeline with stellar temperatures calculated using the direct method of \texttt{isoclassify} \citep{huber2017}. We used $J$ and $K$ photometry available from the EPIC along with the reddening map of \citet{bovy2016} to determine empirical effective temperatures for our sample with the $J-K$ color relation of \citet{gonzalez2009}. We list our effective temperatures calculated with \texttt{isoclassify} in Table \ref{table1}. Our adopted solar reference values are $\nu_{\rm max, \odot}=3090\,$ $\mu$Hz, $\Delta \nu_{\odot}=135.1\, \mu$Hz, and $T_{\rm eff, \odot}=5777$\,K \citep{huber2011}. We calculate our final reported stellar masses and radii using the package \texttt{asfgrid} \citep{sharma16}. As our stars have effective temperatures between 4500 and 5500 K, typical asteroseismic correction factor $f_{\Delta\nu}$ values for all of the stars in our analysis are between 0.98 and 1.02 \citep{sharma16}. We apply this correction factor along with asteroseismic $\nu_\mathrm{max}$ and $\Delta\nu$ values to determine masses and radii in our sample. 

\citet{yu2018} illustrated that fewer than 1\% of asteroseismically confirmed red giant stars smaller than 8 R$_\odot$ have already completed an ascent of the red giant branch, and begun helium burning. Thus, we reject stars larger than 8 R$_\odot$ in order to avoid targeting these ``red clump" stars, which have undergone significantly more evolution than LLRGB stars. We also reject all stars with asteroseismic mass measurement errors greater than 10\% or asteroseismic radius errors larger than 5\%, leaving us with a sample of \nstars\ LLRGB stars with radii between 3.5 and 8 R$_\odot$. We list our asteroseismic frequency parameters and errors and derived asteroseismic masses and radii in Table \ref{table1}. We recover a median radius uncertainty of 3.2\% and mass uncertainty of 5.0\% for our full asteroseismic sample, and 2.2\% in radius and 3.7\% in mass for our LLRGB sample of \nstars\ stars. Figure \ref{starrads} illustrates the distribution of masses and radii for all 5227 asteroseismically vetted stars in our target sample, highlighting our LLRGB star subsample in green.

\begin{figure*}[ht!]
\epsscale{1.15}
\plottwo{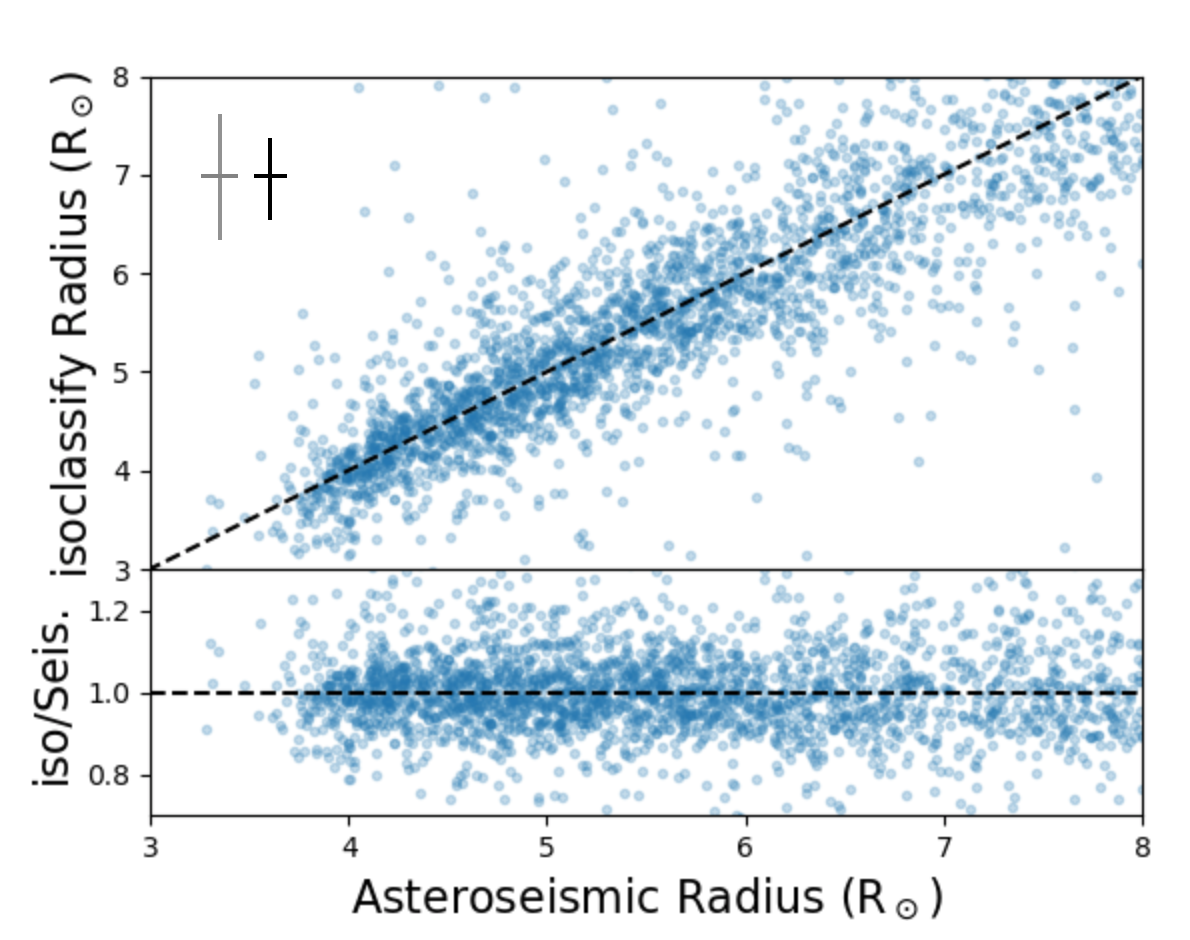}{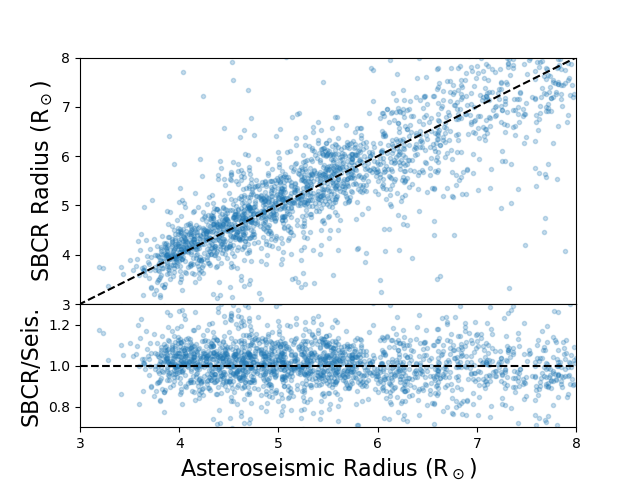}
\caption{Comparison of stellar radii determined through parallax and asteroseismic methods. Left: Radii determined using \texttt{isoclassify} with JHK photometry and \emph{Gaia} DR2 parallaxes are compared against our asteroseismic estimates. Right: A surface brightness--color relation from \citet{graczyk2018} and reddening maps of \citet{bovy2016} have been used to calculate stellar radii, which are then compared against our asteroseismic radii. The scatter in radius determination is larger than the typical offset between parallax-dependent and asteroseismic methods.  Significant outliers from these trends tend to be stars with particularly weak or uncharacteristic asteroseismic signals, unreliable parallax measurements, or at large distances where local reddening can strongly affect photometric temperature estimates. Typical errors for significant outliers and non-outliers are shown in gray and black in the left figure, respectively.}\label{starradcomp}
\end{figure*}

\subsection{Cross-Validation with Independent Radius Estimates}

In order to ensure our asteroseismic results are robust, we use \emph{Gaia} DR2 parallax measurements to determine stellar radii independently and validate our results \citep{gaia2018}. We calculate radii using \emph{Gaia} parallaxes with two different methods. First, we combine JHK magnitudes with the combined reddening map of \citet{bovy2016} to calculate stellar temperatures using the relation of \citet{gonzalez2009} and stellar radii via the Stefan--Boltzmann relation using the \texttt{isoclassify} package  \citep{boltzmann1884,huber2017}. 

We also compute stellar radii for our sample using the surface brightness-color relation of \citet{graczyk2018}. Surface brightness relations are calibrated using directly measured angular diameters from interferometry \citep{kervella2008, boyajian2014} and have been applied to measure precise distances to nearby galaxies many times in the literature \citep[e.g.,][]{kudritzki2014}. For our study, dereddened $V-K$ colors were calculated using the reddening maps of \citet{bovy2016} for 2MASS K magnitudes and CTIO V magnitudes, applied to the 2MASS K and APASS V magnitudes for these stars. These dereddened colors were then converted into angular diameters, which were then combined with \emph{Gaia} parallaxes to determine stellar radii using the relations found in \citet{graczyk2018}. 

Figure \ref{starradcomp} highlights the differences in radius between our asteroseismically determined and parallax-derived radii. We find good agreement between the three sets of stellar radii, with a standard deviation of 10\% for both parallax-driven radius determination methods, and a median offset between asteroseismic radii and radii determined with \emph{Gaia} parallaxes of 3\%. Stars with radii inconsistent between methods at a $>$3-$\sigma$ level tend to be those with particularly weak or uncharacteristic asteroseismic signals or at large distances where local reddening can strongly affect photometric temperature estimates.



\begin{figure*}[ht!]
\epsscale{1.2}
\plotone{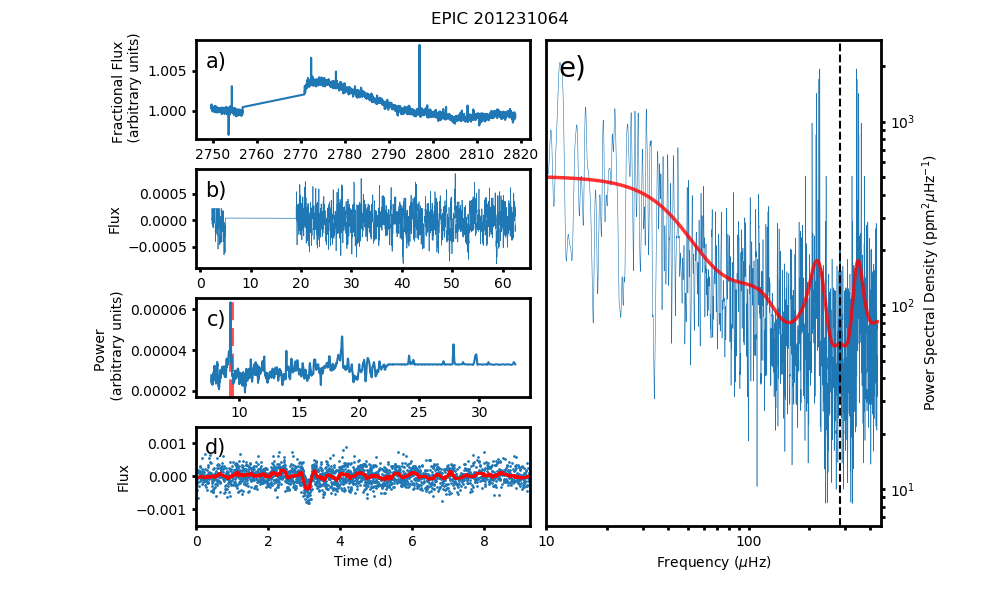}
\caption{Light curve and power spectral density of K2-161, a member of our LLRGB catalog. The raw K2SFF light curve is shown in panel a), whereas a 3-day smoothed version is in panel b) directly underneath. Panel c) displays the power measured by a box least-squares (BLS) search as a function of period, where the red line indicates the best-fit period for that light curve. Panel d) displays the smoothed light curve folded at the best fit period identified by the BLS. The x-axis corresponds to units of time in days for all four panels. Here, the transit of K2-161 b is clearly visible. Panel e), on the right, shows the power spectral density of the smoothed light curve. The dotted vertical line highlights the Nyquist frequency for typical \emph{K2} data. Stellar oscillations are visible above the granulation signal around 220 $\mu$Hz and are mirrored above the Nyquist frequency. A heavily smoothed version of the power spectral density is shown in red, where both the granulation slope and oscillation bump can be seen.\label{lightcurve}}
\end{figure*}

\begin{deluxetable*}{cccccrr}
\tabletypesize{\scriptsize}
\tablecaption{Planets Around LLRGB Stars Found By \emph{K2}\label{table2}}
\tablewidth{0pt}
\tablehead{
\colhead{Name} & \colhead{Planet Radius (R$_\mathrm{J}$)} & \colhead{Orbital Period (days)} & \colhead{Stellar Mass (M$_\odot$)} & \colhead{Stellar Radius (R$_\odot$)} & \colhead{Source}}
\startdata
K2-97b & 1.31 $\pm$ 0.11 & 8.406 $\pm$ 0.0015 & 1.16 $\pm$ 0.12 & 4.2 $\pm$ 0.2 & \citet{grunblatt2017} \\
K2-132b & 1.30 $\pm$ 0.07 & 9.175 $\pm$ 0.0015  & 1.08 $\pm$ 0.12 & 3.8 $\pm$ 0.2 & \citet{grunblatt2017}  \\
K2-161b & 1.45 $^{+0.19}_{-0.14}$\tablenotemark{a} & 9.283 $\pm$ 0.002 & 1.09 $\pm$ 0.10 & 4.12 $\pm$ 0.14\tablenotemark{a} & this work

\enddata 
\tablenotetext{a}{Revised from \citet{mayo2018}.}

\end{deluxetable*}

\section{Planetary Analysis}




\subsection{Planet Sample and Reanalysis of K2-161}

The planets included in our sample are K2-97, K2-132, and K2-161 \citep{grunblatt2016,grunblatt2017,mayo2018}. All three planets are warm ($>$ 150 $F_\oplus$) gas giants larger than Jupiter{\bf\footnote{Here, we have used $F_\oplus$ to represent the incident flux on Earth.}}. For all systems, stellar parameters have been determined through both spectroscopy and asteroseismology as described in the above publications. Planet parameters have been determined through a box least squares search as described in the following subsection and subsequent transit modeling of the \emph{K2} light curves as described in \citet{grunblatt2017} and \citet{mayo2018}. We list the parameters of the planetary systems in Table \ref{table2}. 


\begin{figure}[ht!]
\epsscale{1.15}
\plotone{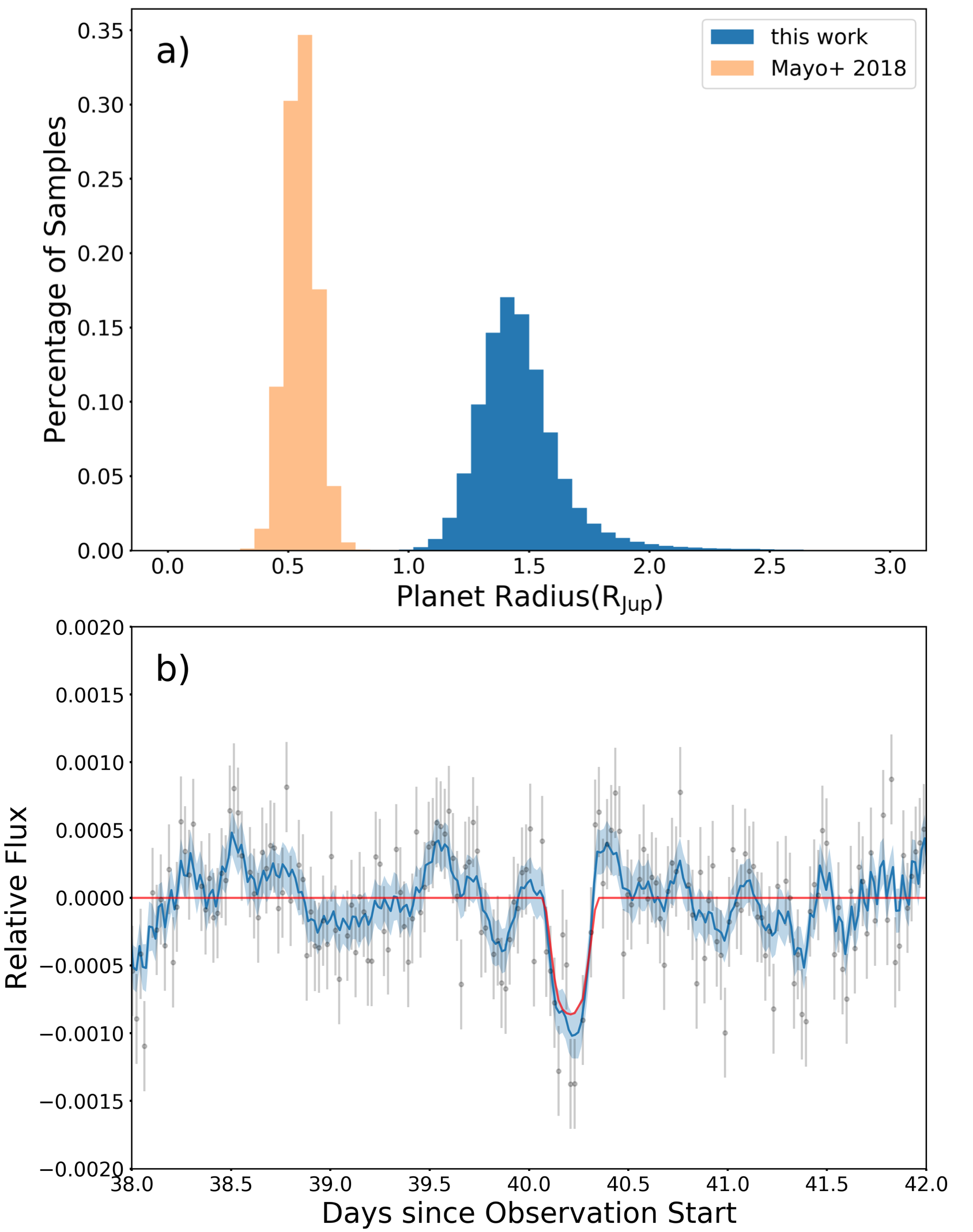}
\caption{Panel a): Planet radius posterior distributions from our analysis of K2-161 b. We find a planet radius that is significantly larger that \citet{mayo2018} due to an increase in our determined stellar radius from both asteroseismology and \emph{Gaia} parallaxes. Panel b): Our combined Gaussian process and transit fit to K2-161, shown in blue. \emph{K2} data is shown in gray, a pure transit fit is given in red. \label{k2161radius}}
\end{figure}

K2-161 b, also known as EPIC 201231064.01, was originally validated by \citet{mayo2018} as a 0.5 $\pm$ 0.1 R$_\mathrm{J}$ planet orbiting a 2.6 $\pm$ 0.3 R$_\odot$ star, with stellar parameters determined by applying the Stellar Parameter Classification (SPC) tool \citep{buchhave2012} to TRES spectra. We note that of all the systems validated by the \citet{mayo2018} study, this particular star had the lowest surface gravity of any validated host star in their sample.

As K2-161 was also a target of our Giants Orbiting Giants \emph{K2} Guest Observer campaign GO10036, we follow the procedure of \citet{grunblatt2017} using asteroseismology to analyze both the light curve and the frequency spectrum of the light curve of this target (see Figure \ref{lightcurve}). We identify an asteroseismic power excess (Figure \ref{lightcurve}e, $\nu_\mathrm{max}$ = 217.6 $\pm$ 0.9 $\mu$Hz, $\Delta\nu$ = 16.80 $\pm$ 0.04 $\mu$Hz), and using equations (4) and (5) determine a stellar mass of 1.09 $\pm$ 0.10 M$_\odot$ and radius of 4.12 $\pm$ 0.14 R$_\odot$. Though the stellar mass determinations between the asteroseismic and spectral analysis are in good agreement (the spectral analysis of \citet{mayo2018} gives a mass of 0.99$^{+0.08}_{-0.06}$ M$_\odot$), the spectroscopic stellar radius (2.6 $\pm$ 0.3 R$_\odot$) and asteroseismic stellar radius (4.12 $\pm$ 0.14 R$_\odot$) determinations disagree at the 3.5-$\sigma$ level. To resolve this discrepancy, we have also determined the radius of K2-161 using a parallax measurement from \emph{Gaia} Data Release 2 and colors using the MIST grid of stellar models via the \texttt{isoclassify} package to ensure that our asteroseismic estimates are robust \citep{choi2016, huber2017}. This parallax-driven stellar grid model analysis gives a stellar radius of 4.2 $\pm$ 0.2 R$_\odot$, in good agreement with the asteroseismic stellar radius determination. Thus, we report the asteroseismic stellar radius here and perform a reanalysis of validated planet K2-161 b using the asteroseismic stellar parameters determined by this work.

\begin{deluxetable}{cccccrr}
\tabletypesize{\scriptsize}
\tablecaption{Planet Transit Model Parameters for the K2-161 System \label{table_k2161}}
\tablewidth{0pt}
\tablehead{
\colhead{Property} & \colhead{Best-Fit Value} & \colhead{Prior} }
\startdata
Period & 9.28316 d & fixed \\ 
Stellar density & 0.022 $^{+0.004}_{-0.005}$ g cm$^{-3}$ & $\mathcal{N}$(0.022; 0.004)\tablenotemark{a} \\
Time of transit & 2752.685 $\pm$ 0.007 d &  (0, $\infty$)\tablenotemark{b} \\
Impact parameter & 0.942 $^{+0.015}_{-0.013}$ & (0, $\infty$) \\
$\nu_\mathrm{max}$ & 222.85 $\pm$ 1.85 $\mu$Hz & (0, $\infty$) \\
$R_p/R_*$ & 0.036 $^{+0.004}_{-0.003}$ &  (0, $\infty$) 
\enddata 
\tablenotetext{a}{Adopted from asteroseismic analysis.}
\tablenotetext{b}{Time given is BJD-2454833.}
\end{deluxetable}

For our reanalysis of the K2-161 b transit signal in the \emph{K2} lightcurve, we follow the analysis method of \citet{grunblatt2017}. We use the \texttt{celerite} package to model the stellar granulation and oscillations seen in the stellar light curve as a sum of periodic and aperiodic simple harmonic oscillator terms, as well as a white noise floor \citep{foreman-mackey2017}. We then use the package \texttt{python-bls} to identify the planet period using a box least-squares analysis \citep{kovacs2002}, and \texttt{ktransit} to model the planet transit and stellar variability simultaneously \citep{barclay2015}. The best fit star and planet model was determined using \texttt{emcee} \citep{foremanmackey2013}. We report the orbital parameters determined through our analysis for the K2-161 system in Table \ref{table_k2161}.

Using this approach and updated stellar parameters, we measure a planet radius of K2-161 b of 1.45 $^{+0.19}_{-0.14}$ R$_\mathrm{J}$ (see Figure \ref{k2161radius}). This is significantly higher than the planet radius reported by \citet{mayo2018} due to two reasons: the larger stellar radius determined by both \emph{Gaia} data and asteroseismology implies that the planet transit duration is uncharacteristically short, which could be due to either a high planet eccentricity or a high impact parameter for the transit, making a larger planet radius more likely. The combination of a larger host star and higher likely impact parameter both imply a larger planet radius, resulting in the discrepancy between planet radii reported here and in \citet{mayo2018}.

Previous high-resolution imaging and spectroscopy shows that there are no widely-separated, bright stellar companions to K2-161 \citep{mayo2018}. To investigate the validation of K2-161 b we obtained six high-precision radial velocity measurements using Keck/HIRES and adaptive optics imaging including aperture masking using Keck/NIRC-2 between January and July 2019. AO imaging revealed a tentative detection of a near equal-mass companion close to K2-161. However, the HIRES spectra show no sign of secondary lines \citep{kolbl2015} and the RV measurements are consistent at the 100 m/s level with complete phase coverage over the planetary orbit, implying that K2-161 b has a planetary mass. Furthermore, the asteroseismic power spectrum shows no signature of a near equal mass companion. Therefore, despite the somewhat ambiguous results of our follow-up observations, we assume that the statistical validation of K2-161 b as a planet holds for the purpose of calculating planet occurrence for our sample. We list the updated stellar and planet parameters for the K2-161 system in Table \ref{table_k2161}. In the case that K2-161 system cannot be confirmed as a {\it bona fide} planet, we address the adjustments to our planet occurrence results in Section 5.1.

\subsection{Planet candidate vetting}

Potential planet transit candidates were identified by comparing the signal detection efficiency of the highest peak given by the BLS search relative to the average signal detection efficiency over all periods searched for a light curve. Signals were deemed to be potential planet signals if the power of the highest peak was between 2 and 4 times the average signal detection efficiency.  This restriction biased our survey against the detection of smaller planets around larger stars, which are difficult to distinguish from noise for these LLRGB stars \citep{sliski2014}. This search returned 52 potential planet candidates, which were then vetted by eye to remove asymmetric or inverted transit features, or those which were too long or short-duration to be likely to be attributable to a planet transit. This left three K2SFF light curves, which were then compared to EVEREST and NASA PDC-MAP light curves \citep{smith2012, luger2016} to ensure the features were astrophysical in origin, and then used to determine final planet parameters (Table \ref{table2}). Once confirming the signals were present in all 3 light curves, the targets were selected for ground-based RV screening.

In addition to the targets identified through our BLS search, additional LLRGB stars in our sample were targeted for RV observations. These stars displayed lower signal-to-background transit-like features that were selected by a visual inspection. However, no planetary RV signals have been identified at orbital periods similar to the transit features seen in these light curves, highlighting the difficulty in detecting planets around more evolved stars \citep{sliski2014}. In the future, these observations will serve as a test of RV stellar variability in LLRGB stars to compare against predictions of such values \citep{tayar2018} and potentially measure precise stellar parameters through the further identification of oscillations \citep{farr2018}.

\subsection{Injection/recovery test}


In order to determine sensitivity to transiting planets in our dataset, we apply the methodology of earlier planet occurrence studies \citep[e.g.,][]{petigura2013, vansluijs2018} to our LLRGB star sample. We injected transit signals from simulated planets on logarithmically uniform, random distributions of periods between 3 and 50 days and linearly uniform planet-to-star radius ratios between 0 and 0.045 into all target light curves with measured asteroseismic signals.  After injecting planet transits into our light curves, we then performed a box least-squares transit search on these light curves \citep{kovacs2002} over the same 3-50 day range in orbital period to see if the transits injected could be recovered. We considered the transit to be recovered if the injected period and the recovered period of the planet agreed to within 0.05 days. Our choice of planet period cut is less stringent than previous studies with \emph{Kepler} \citep[e.g.][]{petigura2013}, due to the shorter time baseline of \emph{K2} campaigns, as well as the high intrinsic variability of evolved stars. This intrinsic variability introduces additional complications in accurately determining the mid-time of transit, limiting orbital period precision for giant stars. 

We validate our transit recovery algorithm by visually inspecting light curves where either injected transits of planets larger than Jupiter on orbits shorter than 10 days were undetected, or planets smaller than 0.5 R$_\mathrm{J}$ on orbits shorter than 10 days were detected. We find that our visual inspection did not recover any of the $>$1 R$_\mathrm{J}$ planets that went undetected at short periods. Thus, we find that our automated planet detection is consistent with our visual planet detection where our transit injection recovery completeness is $>$50\% (see next Section).  Visual inspection also did recover $\sim$60\% of planets smaller than 0.5 R$_\mathrm{J}$ on orbits shorter than 10 days recovered by our algorithm, indicating that occurrence estimates will be reasonably accurate but less precise in regimes where injection recovery completeness is low. We also test our transit recovery algorithm on the light curves without injected transits. Our algorithm successfully detects all three planet transits in our dataset, and does not detect any other false positive planet transits at similar durations and signal to noise ratios.

\subsection{Transit Sensitivity and Survey Completeness}

\begin{figure*}[ht!]
\begin{centering}
\epsscale{1.15}
\plotone{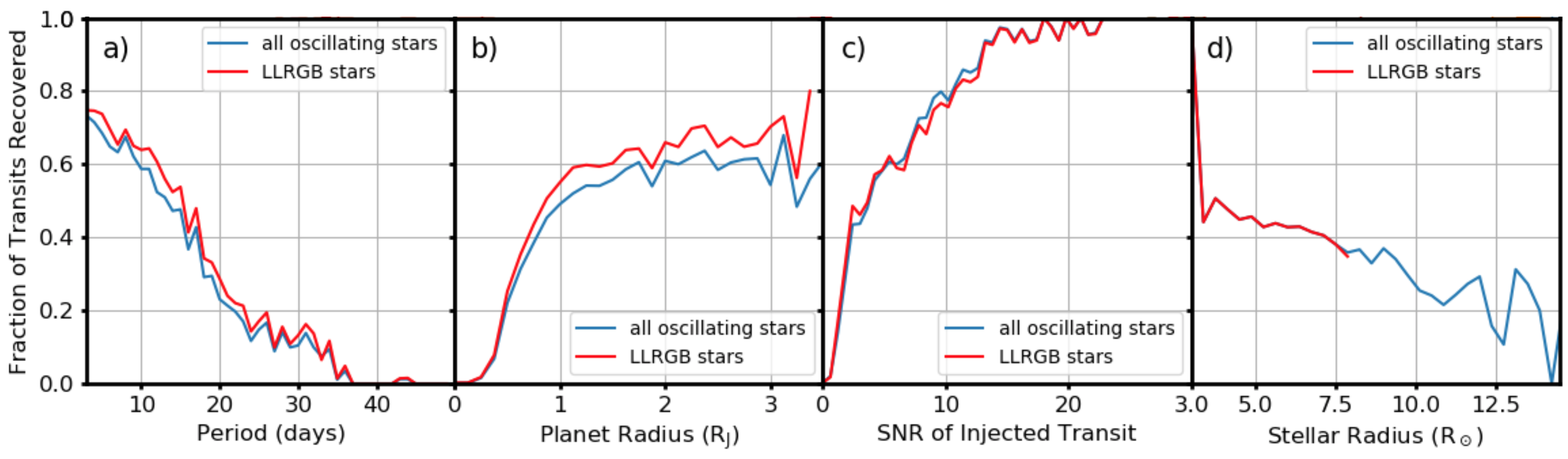}
\caption{Panel a): Sensitivity to injected transit signals as a function of orbital period. Planets are less detectable at longer orbital periods around all stars. Panel b): Sensitivity to injected transit signals as a function of planet radius. Using only stars smaller than 8 R$_\odot$, more than 60\% of planets larger than Jupiter are detected. Panel c): Fraction of injected transits recovered by our pipeline, as a function of injected transit signal to noise ratio in the light curve. Restricting our sample to only LLRGB stars has no significant effect on our results. Panel d): Sensitivity to injected transit signals as a function of stellar radius. Stellar radii have been determined through asteroseismology. Above 8 R$_\odot$, planet transit detectability drops below 40\%. \label{mes}}
\end{centering}
\end{figure*}

Figure \ref{mes} plots the our ability to recover signals injected into our light curves. We compare results of this injection/recovery test for our full sample of vetted asteroseismic stars to the subset of LLRGB stars identified by our study. We show (from left to right) the fraction of injected signals recovered as a function of orbital period, planet radius, transit signal to noise ratio, and stellar radius. 

Figure \ref{mes}a shows the fraction of injected signals recovered in our dataset as a function of orbital period. Sensitivity decreases as a function of orbital period, as fewer transits can be identified in a single 80-day \emph{K2} campaign. Little variation between the LLRGB stars and the full asteroseismic sample is seen.

The fraction of injected transits recovered as a function of planet radius is shown in Figure \ref{mes}b. Sensitivity increases with planet radius, reaching a plateau of about 60\% by a radius of  1.5 R$_\mathrm{J}$. This detection plateau feature may be due to intrinsic faintness of certain stars, systematic variability that is simply too large to allow any planet transit to be detected through a straightforward box least-squares search, or the limited detection opportunities for planets injected on long period orbits that have only one or two transits in a \emph{K2} campaign.

Figure \ref{mes}c displays the fraction of transits detected as a function of signal to noise ratio in our sample. We can detect more than half of planet transits at signal to noise ratios of 5 or better, and more than 95\% at signal to noise rations above 16, in agreement with previous \emph{Kepler} occurrence studies \citep{fressin2013}. Thus we expect that $>$50\% of planets with a transit signal to noise of 5 or better and an orbital period less than 50 days have been detected in our sample, in good agreement with our findings in Figure \ref{mes}b. 

However, recovery of transit signals for a given planet radius does depend on the stellar radius. Figure \ref{mes}d illustrates the fraction of injected signals recovered in our dataset as a function of stellar radius. We can see that detectability decreases with stellar size. Below $\sim$4 R$_\odot$ and above $\sim$10 R$_\odot$, our stellar sample becomes too small for reliable statistics, resulting in imprecise estimates of transit recovery.


\begin{figure*}[ht!]
\epsscale{1.15}
\plottwo{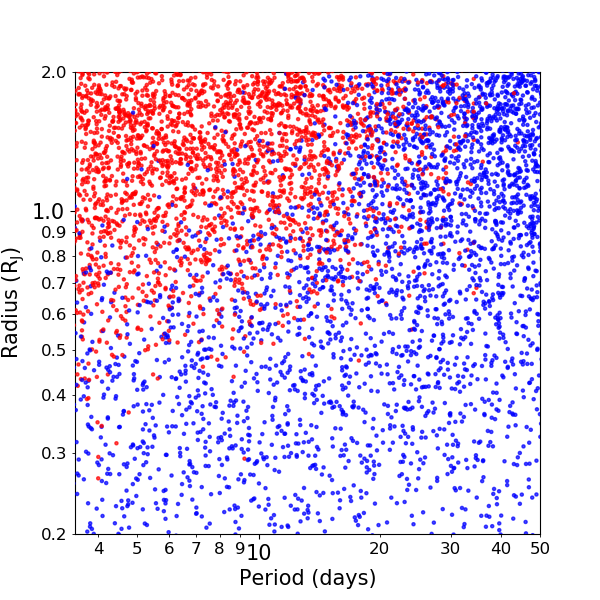}{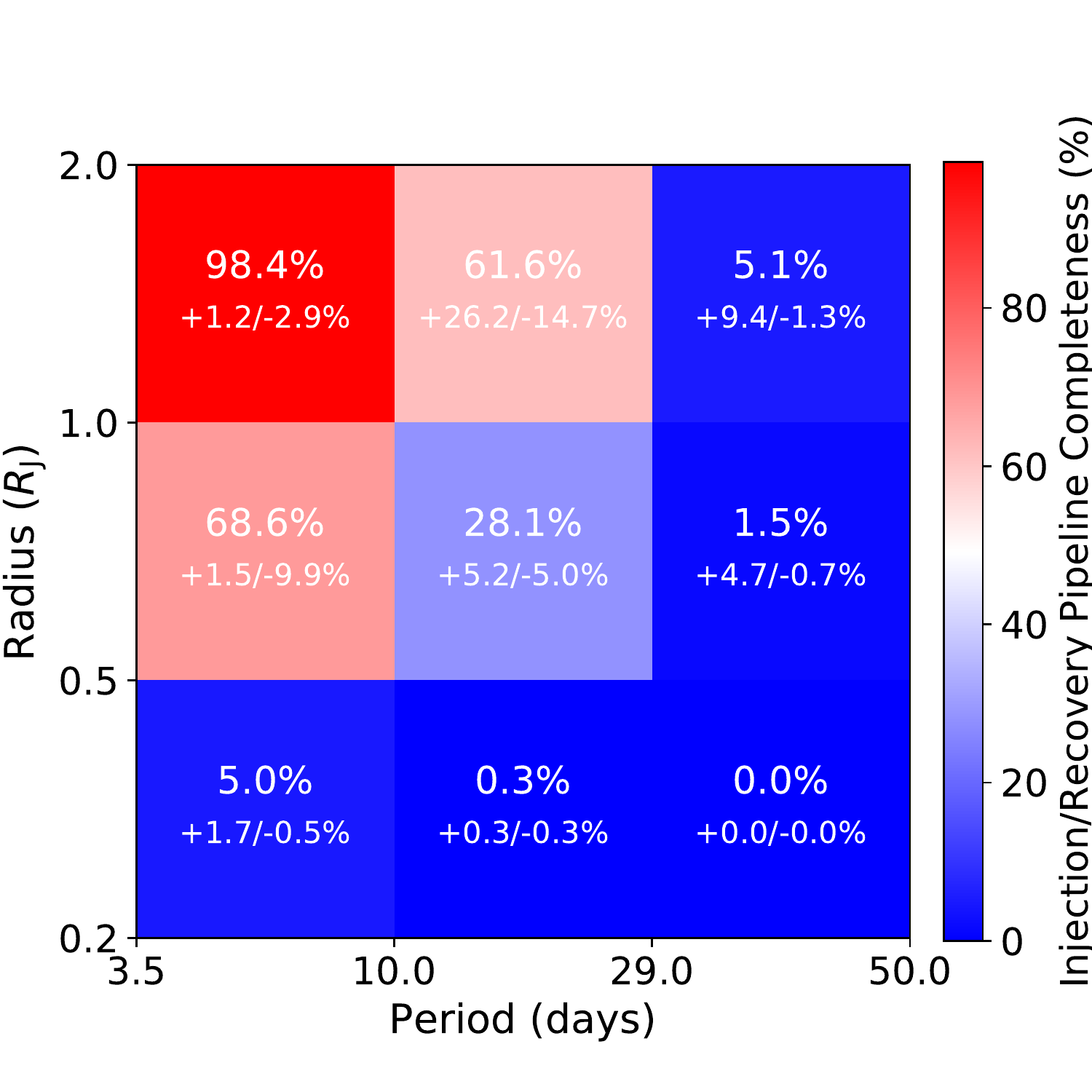}
\caption{Left: Transits injected into our LLRGB light curves, as a function of planet radius and orbital period. Transits that were recovered are shown in red, while those missed are shown in blue. Right: Injection/recovery survey completeness for our sample of oscillating, $<$8 R$_\odot$ stars. We see that the survey is largely complete for planets larger than 1 R$_\mathrm{J}$ on orbital periods shorter than 10 days, and are more than 50\% complete for planets down to 0.5 R$_\mathrm{J}$ as well as super-Jupiter sized planets out to 25 day orbital periods. 
\label{completeness}}
\end{figure*}

In order to evaluate our survey completeness and calculate the occurrence of planets around these stars, we need to understand the properties of planets that could have been detected by our survey, and compare it to the planets we actually found. The left panel of Figure \ref{completeness} illustrates the distribution of transits as a function of planet radius and period injected in our transit injection and recovery test. Injected signals that were recovered are shown in red, while those that went undetected by the box least-squares search are shown in blue. We inject transits around all \nstars\ LLRGB stars in our sample.

We then evaluate our completeness fraction in bins of planet radius and orbital period, with upper and lower uncertainties estimated by calculating completeness with a period precision threshold of our injection/recovery pipeline set to 0.1 and 0.03 day precision, respectively. For our completeness estimate, we require our recovered period to agree with the injected period by 0.05 days, comparable to the period precision required by earlier \emph{Kepler} transit injection/recovery tests \citep{petigura2013}. Completeness may also vary within the bins in orbital period and planet radius specified here. The right panel of Figure \ref{completeness} illustrates the completeness of our survey. We find that we are sensitive to almost all planets at periods between 3.5 and 10 days and larger than Jupiter, with sensitivity dropping at larger periods and smaller planet radii, reaching less than 50\% completeness at periods greater than 29 days and planet radii smaller than 0.5 R$_\mathrm{J}$. 




\subsection{Planet Occurrence Calculation}

\begin{figure}[ht!]
\epsscale{1.2}
\plotone{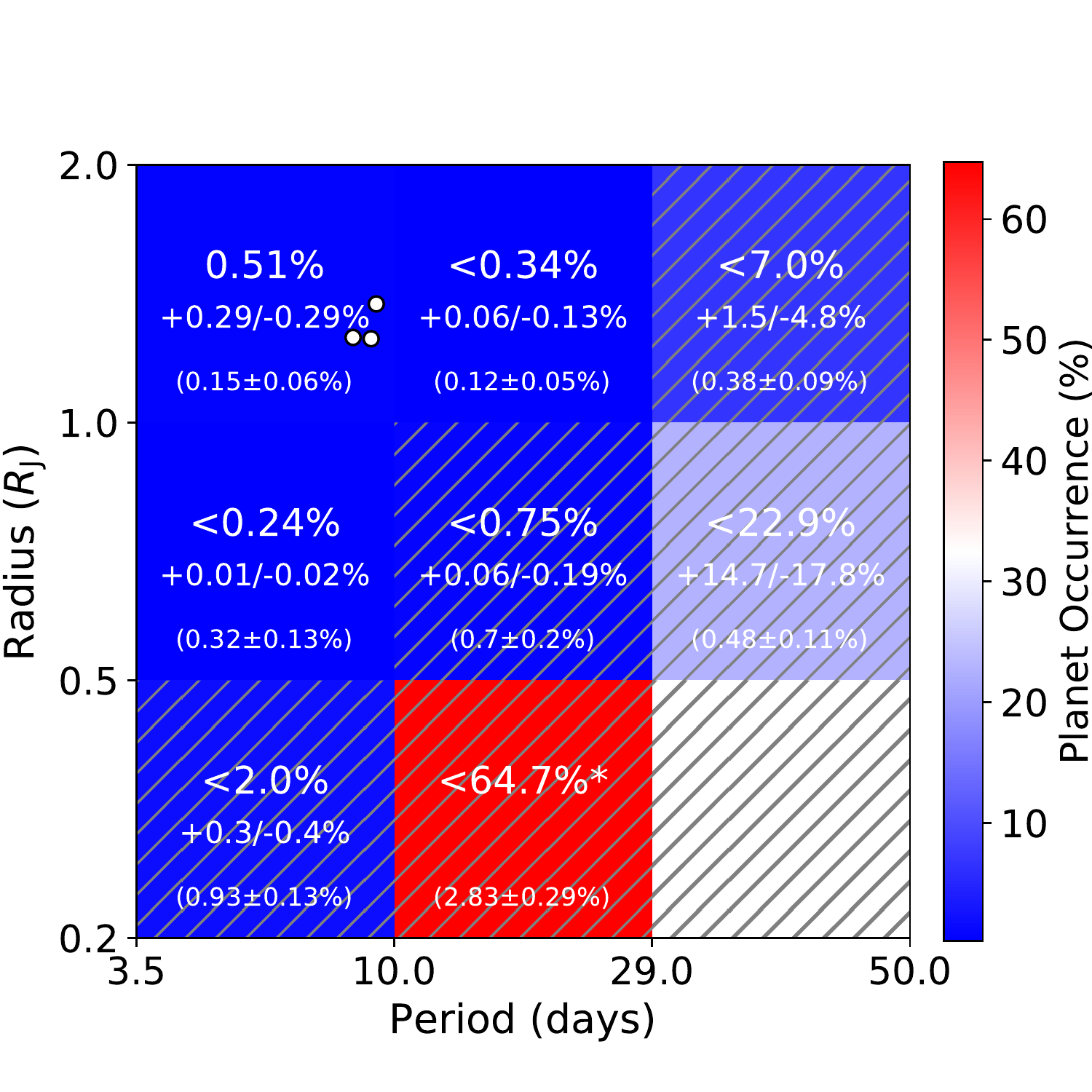}
\caption{Planet occurrence around 3.5-8 R$_\odot$ stars observed by \emph{K2}, as a function of orbital period and radius. In those bins where no planets were found, upper limits were calculated for planet occurrence. Hatched cells designate where injection/recovery completeness is below 50\% for our stellar sample, and asterisk indicates where uncertainties on completeness was too large to be reliably estimated. Main sequence occurrence rates from \citet{howard2012} are shown in parentheses at the bottom of each bin. Planets detected in this survey are shown by the white markers. For planets with radii larger than Jupiter at orbital periods less than 10 days, we find a consistent yet tentatively higher number of planets orbiting our sample of LLRGB stars than main sequence stars. For all regions of parameter space where planets were not found, the upper limits of planet occurrence calculated by this survey are in agreement with the main sequence occurrence rates reported by \citet{howard2012}.
\label{occurrence}}
\end{figure}

\begin{figure*}[ht!]
\epsscale{1.2}
\plotone{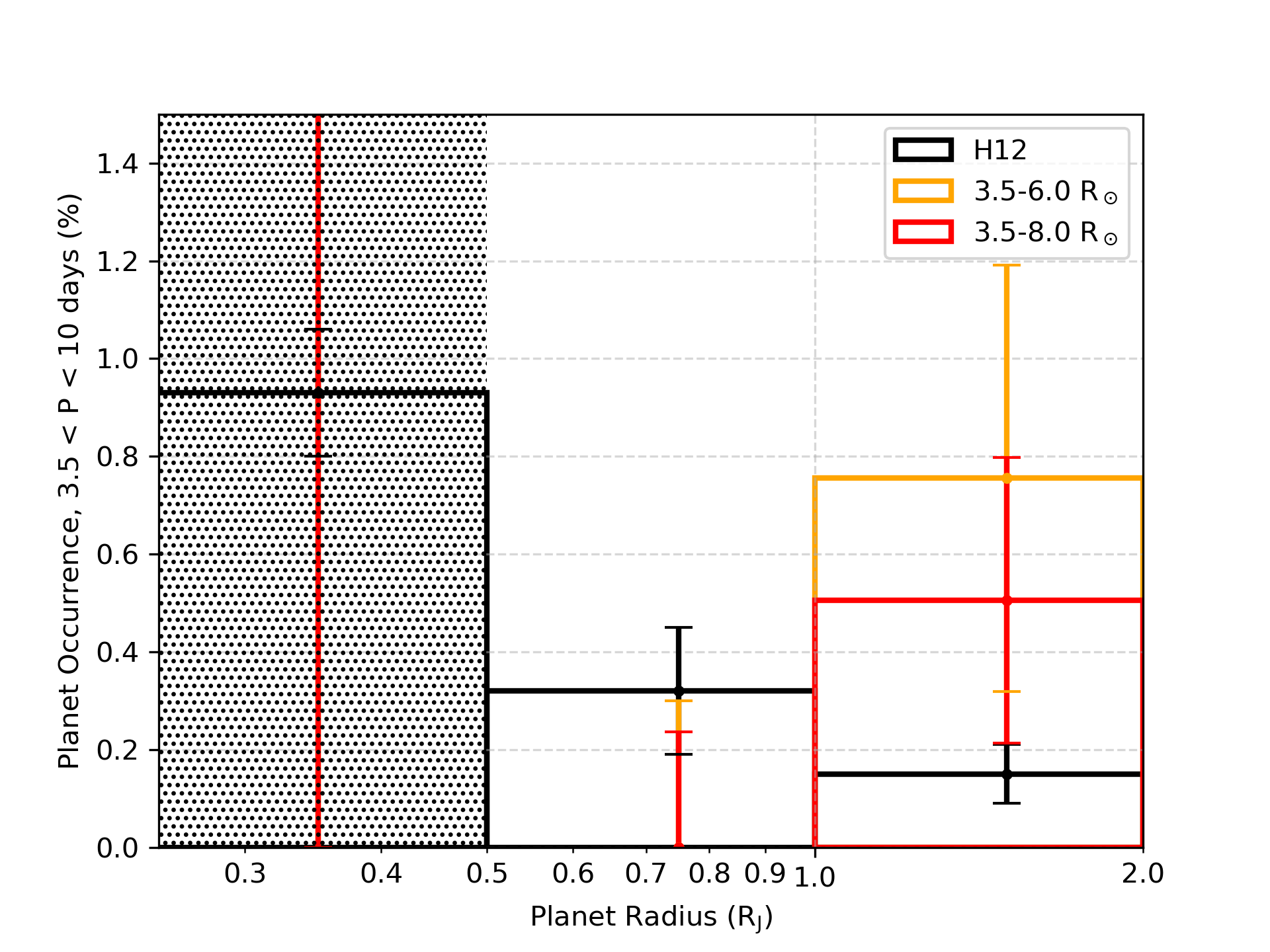}
\caption{Planet occurrence within 3.5-10 days as a function of planet radius, with different stellar radius  populations broken into groups. Planet radius is given on the x-axis, while different stellar populations separated by radius are indicated in the plot legend in the upper right. Errors indicate 68\% confidence intervals. The shaded region indicates where our transit injection/recovery completeness was below 50\%. Planet occurrence appears consistent yet tentatively higher around evolved stars with radii between 3.5 and 8 R$_\odot$ than main sequence stars for planets larger than Jupiter. This difference seems enhanced when only evolved stars between 3.5 and 6 R$_\odot$ are considered. Close-in planet occurrence may be similar or lower around evolved stars than main sequence stars at planet radii between 0.5 and 1 R$_\mathrm{J}$.
\label{radocc}}
\end{figure*}

In order to calculate planet occurrence, we followed the prescription of \citet{howard2012}, using our survey completeness to estimate how many planets we could have found, and then compare that to the number of planets we actually found in each bin. 

For each orbital period/planet radius ($P$--$R_P$) bin, we count the number of transiting planets, $n_\mathrm{pl,bin}$. 
As we assume planetary orbits to be randomly oriented, each transiting planet represents a larger number of planets that are not transiting. We compute this augmented number of planets, 
\begin{equation}
n_\mathrm{pl,aug,bin} = \sum_{n=0}^{i}a_i/R_i, 
\end{equation}

where $i$ is the number of planets per bin, $a_i$ is the semimajor axis of a given planet $i$, and $R_i$ is the stellar host radius, to account for non-transiting planets. We note that this overestimates the detection efficiency, and underestimates the occurrence in our case, if the intrinsic detection efficiency changes significantly across the finite width of the bin in both orbital period and planet radius \citep{hsu2018}. However, we use this method to allow for direct comparison to the results of \citet{howard2012}, which were computed using the same inverse detection efficiency method.


The planets considered by our survey and their physical properties are listed in Table \ref{table2}. To compute occurrence, we divide the number of stars with detected transiting planets in a particular bin by the number of stars around which a transiting planet could have been detected in a given bin, n$_\mathrm{*,amen}$. This number is just our total number of LLRGB stars, N = \nstars\,, multiplied by the completeness in a bin computed in our injection/recovery test. The debiased fraction of stars with planets per $P$--$R_p$ bin, $f_\mathrm{bin}$, is given by
\begin{equation}
f_\mathrm{bin} = n_\mathrm{pl,aug,bin}/n_{*,\mathrm{amen}}.
\end{equation}

For those bins where no planets were found, we place an upper limit on planet occurrence by calculating the planet occurrence if one planet had been detected in that bin. In those bins where we have detected planets, we find that our errors in completeness are negligible compared to Poisson errors introduced by our small sample of detected planets, which dominate our errors in planet occurrence. Errors on occurrence upper limits are calculated by propagating our errors in survey completeness forward, and adding Poisson errors in quadrature where planets were detected. Poisson errors dominate where planets were detected around both main sequence and LLRGB stars.

Figure \ref{occurrence} shows planet occurrence in our sample. Our occurrence estimate and errors are shown at the center of each bin, while main sequence occurrence rates determined from \citet{howard2012} for that bin are given in parentheses below our estimates. In order to test our sensitivity to stellar radius, we also repeat our experiment, excluding all stars with asteroseismic radii larger than 6 R$_\odot$, a subsample of \nsmallstars\ stars. The results of this test, along with all of our other occurrence estimates, are shown in Table \ref{table:planocc}.


\section{Discussion}

\subsection{Close-In Giant Planet Occurrence of Evolved Stars}

Based on our analysis of \nstars\ oscillating stars, we find that 0.49\% $\pm$ 0.28\% of stars with radii between 3.5 and 8 R$_\odot$ host planets larger than 1 R$_\mathrm{J}$ on 3.5--10 day orbits, a consistent yet tentatively larger fraction than the 0.15\% $\pm$ 0.06\% of main sequence stars found to be hosting similar planets by \citet{howard2012}. We also find that fewer than 0.23\%$^{+0.02\%}_{-0.01\%}$ of LLRGB stars host planets with radii between 0.5 and 1 R$_\mathrm{J}$ and orbital periods of 3.5--10 days, in agreement with the 0.32\% $\pm$ 0.13\% of systems found around main sequence \emph{Kepler} stars. Our upper limit of 0.33\%$^{+0.07\%}_{-0.12\%}$ of stars hosting planets $>$ 1 R$_\mathrm{J}$ on 10--29 day orbital periods is also in agreement with the 0.12\% $\pm$ 0.05\% of stars found to be hosting such planets on the main sequence \citep{howard2012}. Due to the intrinsic variability of giant stars and the limited duration of \emph{K2} time-series, we are not sensitive to transit signals from planets at orbital periods of 29 days or longer, or 0.5 Jupiter radii or smaller, or for planets between 0.5 and 1.0 R$_\mathrm{J}$ on 10 to 29 day orbital periods in our sample. 


If we restrict ourselves to stars smaller than 6 R$_\odot$, we find that occurrence is enhanced at the smallest orbital periods and largest planet radii, as all planets considered in our survey orbit relatively small ($\sim$4 R$_\odot$) stars in our sample at periods $<$10 days. We find that 0.72\% $\pm$ 0.41\% of stars with radii between 3.5 and 6 R$_\odot$ host planets larger than 1 R$_\mathrm{J}$ on 3.5--10 day orbits. Upper limits on planet occurrence are particularly large in the longer period bins for these less evolved stars, due to the smaller sample size. At orbital periods greater than 10 days, our maximum planet occurrence estimates are in complete agreement with main sequence planet occurrence rates \citep{howard2012, fressin2013}. We state these planet occurrence findings in Figure \ref{radocc} and Table \ref{table:planocc}.

We note that current ambiguities in the observations of K2-161 may prevent the confirmation of K2-161b as a planet, and may even indicate that the planet signal is a false positive (Section 4.1). This would result in a reduction of the occurrence rates of planets larger than Jupiter at orbital periods less than 10 days to 0.33\% around LLRGB stars. However, since this change falls within the reported error bars for planet occurrence, we find that the ambiguous nature of this validated planet does not significantly affect the interpretation of our results, and thus treat K2-161b as a confirmed planet for the purposes of this analysis.

\begin{deluxetable*}{cccccccc} 
\tablecolumns{6} 
\tablewidth{0pt} 
\tablecaption{Comparison of Planet Occurrence Around Main Sequence and Evolved Stars \label{table:planocc}} 
\tablehead{ 
 & & & & \colhead{Planet Period}& & \\
\colhead{Planet Radius} & \colhead{Planet Sample} &\colhead{Stellar Sample} & & & & \\
& & & \colhead{3.5-10 days} & \colhead{10-29 days} &  \colhead{29-50 days} \\
}
\startdata 
 & Main Sequence & Main Sequence & 0.15 $\pm$ 0.06 & 0.12 $\pm$ 0.05 & 0.38 $\pm$ 0.09\\
 & Main Sequence & LLRGB & 0.28 $\pm$ 0.16 & 0.27 $\pm$ 0.16 & \tablenotemark{*}\\
\vspace{-0.2cm} 1-2 R$_\mathrm{J}$ & & & & & \\
& LLRGB & LLRGB & 0.51 $\pm$ 0.29 & $<$0.34$^{+0.06}_{-0.13}$ & $<$7.0$^{+1.5}_{-4.8}$\tablenotemark{*}\\
 & LLRGB ($<$6 R$_\odot$) & LLRGB ($<$6 R$_\odot$) & 0.76 $\pm$ 0.44 & $<$0.48$^{+0.12}_{-0.28}$ & $<$10.7$^{+2.3}_{-8.7}$\tablenotemark{*} \vspace{0.1cm}\\
 \hline \vspace{-0.1cm}\\ 
 & Main Sequence & Main Sequence & 0.32 $\pm$ 0.13 & 0.70 $\pm$ 0.20 &  0.48 $\pm$ 0.11\\
 & Main Sequence & LLRGB & 0.27 $\pm$ 0.12 & 0.35 $\pm$ 0.04\tablenotemark{*}& \tablenotemark{*}\\
 \vspace{-0.2cm} 0.5-1 R$_\mathrm{J}$ & & & & & \\
 & LLRGB & LLRGB & $<$0.24$^{+0.01}_{-0.02}$ & $<$0.75$^{+0.06}_{-0.19}$\tablenotemark{*} & $<$22.9$^{+14.7}_{-17.8}$\tablenotemark{*} \\
 & LLRGB ($<$6 R$_\odot$) & LLRGB ($<$6 R$_\odot$) & $<$0.30 $\pm$ 0.11 & $<$0.91$^{+0.34}_{-0.51}$\tablenotemark{*} & $<$25.2$^{+3.8}_{-21.5}$\tablenotemark{*} \vspace{0.1cm}\\
 \hline\\
 & Main Sequence & Main Sequence & 0.93 $\pm$ 0.13 & 2.83 $\pm$ 0.29 & 1.85 $\pm$ 0.41 \\
 & Main Sequence & LLRGB & 23.4 $\pm$ 16.6\tablenotemark{*} & \tablenotemark{*} & \tablenotemark{*}\\
 \vspace{-0.2cm} 0.2-0.5 R$_\mathrm{J}$ & & & & & \\
 & LLRGB & LLRGB & $<$2.0$^{+0.3}_{-0.4}$\tablenotemark{*} & $<$64.7\tablenotemark{*} & \tablenotemark{*} \\
 & LLRGB ($<$6 R$_\odot$) & LLRGB ($<$6 R$_\odot$) & $<$2.4$^{+0.6}_{-1.2}$\tablenotemark{*} & $<$76.7\tablenotemark{*} & \tablenotemark{*}

\enddata 
\tablenotetext{*}{Injection/recovery tests indicate a completeness below 50\% for these regimes. No value is reported in those regimes where no injected signal was recovered.}
\tablecomments{All occurrence values quoted are percentages. Main sequence planets orbiting main sequence star results are taken from \citet{howard2012}.}

\end{deluxetable*}

\subsection{Reproducing Main Sequence Occurrence Rates with Our Pipeline}

We confirm our results are robust by injecting a main sequence population of planets selected using the same constraints as \citet{howard2012} around our targets. Specifically, we inject only confirmed planets with \emph{Kepler} Object of Interest designations below 1650 around stars with  T$_\mathrm{eff}$ = 4100--6100 K, log $g$ = 4.0--4.9 dex, and \emph{Kepler} magnitude K$_p$ $<$ 15 mag around a random subset of our LLRGB star sample.

We take this population of detected transiting planets around main sequence stars and use it to infer the true population of planets around main sequence stars. We do this by computing the inverse detection efficiency for each main sequence transiting planet detection to infer the total population, or augmented number of planets orbiting main sequence stars, equivalent to computing $n_\mathrm{pl,aug}$ for all planets in the main sequence sample. Given a main sequence sample of 58,041 stars \citep{howard2012}, we use this to determine the number of planets per main sequence star, and then compute the expected number of planets orbiting our sample of LLRGB stars, assuming the main sequence and LLRGB planet populations are equivalent. We then inject this population of planets around our LLRGB stars, and use this injected planet population to determine planet occurrence if the main sequence and LLRGB planet populations are equivalent, accounting for transit detection bias differences between the main sequence and LLRGB stellar populations. We then compare this result to the planet occurrence determined by our K2 survey of LLRGB stars. We show our planet occurrence estimates from injecting the main sequence planet population around main sequence stars in Table \ref{table:planocc}.



We find that when placing the main sequence population of planets around LLRGB stars, transiting planets are detected around 0.4\% of stars, comparable to the number of planets observed around main sequence stars where our survey completeness is greater than 50\%. We find that our planet occurrence estimates generated using the reproduced \citet{howard2012} planet sample agree within 1-$\sigma$ with both our observed LLRGB planet population as well as the planet occurrence rates stated in \citet{howard2012} in all bins where our LLRGB survey completeness is greater than 50\%, including the bin where planets were detected in our \emph{K2} survey. We measure a main sequence occurrence rate of 0.28 $\pm$ 0.16\% for planets larger than Jupiter on 3.5--10 day orbits, 0.27 $\pm$ 0.16\% for planets larger than Jupiter on 10--29 day orbits, and 0.27 $\pm$ 0.12\% for planets between 0.5 and 1.0 R$_\mathrm{J}$ on 3.5--10 day orbits, in good agreement with the \citet{howard2012} results.

\subsection{Effects of Stellar Mass and Metallicity}

It is important to account for the variation in stellar masses and metallicities in our sample to ensure that differences in occurrence are a result solely of evolution and not an effect of analyzing different stellar populations. \citet{johnson2010} show that planet occurrence is proportional to stellar mass ($f \propto M$) and has a dependence on stellar metallicity as well ($f \propto 10^{1.2\mathrm{[Fe/H]}}$). The median mass and metallicity of the stars around which we did find planets is 1.09 M$_\odot$ and 0.1 dex higher than solar, respectively. The median mass of our entire sample is  1.18 M$_\odot$. We do not have metallicity estimates for our entire stellar sample, and thus assume the average metallicity of this sample is within 0.1 dex of solar metallicity. Based on these measurements and assumptions, we expect at most a $\approx$30\% higher occurrence rate for our sample than that of truly solar-like stars. We find that when compared to the \citet{howard2012} sample of planet hosts, the median and mean masses of both stellar samples agree to within less than the mass dispersion in either sample, and the mean and median metallicities of our planet host sample are within the dispersion of the \citet{howard2012} main sequence sample \citep{johnson2017}. Thus, the effects of stellar metallicity and mass may not be sufficient to fully account for the differences in planet occurrence found here, but a larger population of planets around evolved stars is needed to definitively distinguish between metallicity and evolutionary effects.

\subsection{Constraints on Planet Dynamics}

It is assumed that the evolution of a host star will strongly affect the orbital and atmospheric properties of any planets in orbit, which may explain the tentative enhancement in short period planets larger than Jupiter seen in this study \citep{villaver2009, lopez2016, grunblatt2018}. Using the planet occurrence distributions of both the main sequence and evolved stars, and assuming a uniform fractional radius inflation for all planets from the main sequence to the evolved stage, we can predict the observed change in planet occurrence as a function of planet radius. This will allow distinction between a static orbit scenario, where planets are inflated by stellar evolution but do not migrate, and a scenario where stellar evolution also causes orbital motion of its planets.

Using the main sequence planet sample of \citet{howard2012}, we assume a monotonic increase in radius of 50\% for the 3.5-10 day orbital period, $>$ 0.5 R$_\mathrm{J}$ main sequence planet population from planet inflation due to stellar evolution. This would increase the occurrence from 0.15\% to 0.21\% of $>$ 1 R$_\mathrm{J}$ planets at orbital periods of 3.5-10 days around evolved stars. This occurrence rate is smaller than that measured for evolved systems by this work and only marginally higher than the observed main sequence occurrence rate. We conclude that inflation of the main sequence population alone is not sufficient to explain the elevated occurrence around evolved stars. Instead, the observed evolved system, short-period super-Jupiters are likely both smaller and at larger orbital distances around main sequence stars.

During post-main sequence stellar evolution, planets on eccentric orbits are likely circularizing while spiraling into their host stars \citep{villaver2009,villaver2014}. This process will presumably cause significant tidal distortion and potential dissipation within the planets, heating their interiors and inflating them to larger radii, producing a population of transient, moderately eccentric close-in planets falling into their host stars \citep{bodenheimer2001,grunblatt2018}. The increased irradiation of these planets by their host stars will have a similar effect \citep{lopez2016,grunblatt2017}. Given that a population of cold, gas giant planets exists around a higher fraction of main sequence stars than close-in planets around evolved stars \citep{bryan2016, ghezzi2018}, inspiral of some or all of the main sequence giant planet population could result in the close-in giant planet population found here. Furthermore, the increase in irradiation may result in photoevaporation of less massive planetary atmospheres, leaving behind undetectable rocky cores alongside the inflated planets we can detect \citep{owen2017}.




The timescale of inspiral of an eccentric, gaseous planet may be inferred from these observations. Following the reasoning of \citet{grunblatt2018}, if we assume a``constant phase lag" model for tidal evolution \citep{goldreich1966,patra2017}, we calculate the inspiral timescale of the planet as 

\begin{equation}
\tau = \frac{2Q_*}{27\pi}\Big(\frac{M_*}{M_p}\Big)\Big(\frac{a}{R_*}\Big)^5 P,
\end{equation}

where $M_*$, $R_*$ and $Q_*$ are the mass, radius, and tidal quality factor of the star, $M_p$ is the planet mass, $a$ is the semimajor axis of the planet's orbit, and $P$ is the planet period. Assuming a stellar tidal quality factor $Q_\star \sim 10^6$ as found in earlier studies \citep{essick2016}, we find an inspiral timescale of $\lesssim$ 2 Gyr for all planets in our sample, potentially much longer than the LLRGB phase of stellar evolution. 

However, the relatively high planet occurrence of the largest, shortest period planets around the smallest stars in our sample that is not seen for larger stars suggests that planetary systems can survive the subgiant phase of stellar evolution, and are being reshaped during this low-luminosity ascent of the red giant branch. A pile-up of planets at small orbital periods around $\sim$4 R$_\odot$ stars would imply that our inspiral timescale for these systems is overestimated, close-in planets only survive past this stellar radius size in rare cases, and may be engulfed soon after the star reaches this size. The strong dependence of inspiral timescale on stellar radius may reflect this change in inspiral speed at larger stellar radii. However, the small sample of planets found combined with a selection effect which strongly favors smaller stars may also be responsible for the stellar radius distribution we observe. Thus, additional systems must be observed to improve models of planet evolution significantly.

\section{Conclusions}

We have identified \nstars\ low-luminosity red giant branch (LLRGB) stars observed in 15 of the first 16 campaigns of the \emph{K2} mission using parallaxes and asteroseismology to determine stellar radii and masses. We then perform a transit injection/recovery test to determine the transit survey completeness, and thus infer planet occurrence around these evolved stars. We find that:

\begin{itemize}

\item Using asteroseismology, we constrain masses and radii of \nstars\ LLRGB stars to 3.7\% and 2.2\% median uncertainties, respectively. Asteroseismic radii for LLRGB stars agree with radii calculated using \emph{Gaia} parallaxes with both surface brightness-color relations and stellar grid modeling with a median offset of 3\% and a scatter of 10\%.

\item At radii larger than 1 R$_\mathrm{J}$ and orbital periods 3.5-10 days, when compared to the main sequence population, planet occurrence appears tentatively higher around evolved stars, yet agrees with main sequence occurrence rates within errors. At orbital periods of 3.5-50 days and planet radii between 0.2 and 2 times the size of Jupiter, upper limits on planet occurrence around evolved stars are in agreement with planet occurrence determined around main sequence stars \citep{howard2012, fressin2013}. 

\item As all confirmed planet hosts in our sample are larger than 3.6 R$_\odot$, planetary systems can survive the subgiant phase of stellar evolution at least until the host star reaches the base of the red giant branch. As no planetary hosts in our sample are larger than 4.4 R$_\odot$, this implies that planetary systems are likely destroyed by their host stars during the early stages of ascent up the red giant branch. Planet occurrence likely varies as a function of stellar radius in our sample.

\item Assuming a 50\% increase in radius of all planets orbiting main sequence stars is insufficient to explain the larger fraction of short-period super-Jupiter sized planets. This suggests that if there is in fact a larger fraction of short-period super-Jupiter sized planets around evolved stars, orbital migration may contribute to this planet population. The difference in stellar mass distribution of our stars relative to main sequence stars is insufficient to account for the difference in planet distribution. The influence of metallicity is more unclear, as metallicities are only known precisely for the planet hosts in our sample, which are marginally more metal-rich than the main sequence planet host population.\\


\end{itemize}

Differences between the occurrence of planets around main sequence and evolved stars gives us valuable information about the evolution of planetary systems in conjunction with the evolution of their host star. However, these results rest on only three planet detections among \nstars\ stars, a sample less than 2\% the size of \emph{Kepler} main sequence occurrence studies, and thus a larger sample of stars and planets will be essential to determining whether these deviations in planet occurrence are significant. 

The recently launched NASA \emph{TESS} Mission is essential to further investigations of the planet population around evolved stars. \emph{TESS} will observe two orders of magnitude more evolved stars than \emph{K2}, and will cover more than 90\% of the sky with a cadence, precision and depth sufficient to identify giant planets orbiting evolved, oscillating stars with orbital periods of 10 days or less at distances $\geq$ 1 kpc  \citep{campante2016, campante2018}. Indeed, the first transiting planet orbiting an oscillating, evolved host star observed by \emph{TESS} has already been found \citep{huber2019}. A survey of the $\sim$400,000 evolved stars observed by \emph{TESS} will thus be instrumental in determining precisely what the fraction of 0.5-1.0 R$_\mathrm{J}$ planets is around evolved stars, and how depleted the population is relative to main sequence stars. Furthermore, due to Malmquist bias evolved stars tend to be further away than main sequence stars of similar magnitudes. Thus, a larger sample of transiting planets orbiting evolved stars could reveal deviations in planet occurrence as a function of location in the Galaxy out to kiloparsec distances. 

\acknowledgements{The authors thank Andrew Mayo, Vincent Van Eylen, Joshua Winn, Songhu Wang, Rolf-Peter Kudritzki, and the California Planet Search team for helpful discussions, and thank Adam Kraus, Aaron Rizzuto, Trent Dupuy and the California Planet Search team for help with initial follow-up observations for K2-161b. S.G. and D.H. acknowledge support by the National Aeronautics and Space Administration under Grant 80NSSC19K0110 issued through the K2 Guest Observer Program. J.~Z. acknowledges support from NASA grants 80NSSC18K0391 and NNX17AJ40G. This research has made use of the NASA Exoplanet Archive, which is operated by the California Institute of Technology, under contract with the National Aeronautics and Space Administration under the Exoplanet Exploration Program. This work was based on observations at the W. M. Keck Observatory granted by the University of Hawaii, the University of California, and the California Institute of Technology. We thank the observers who contributed to the measurements reported here and acknowledge the efforts of the Keck Observatory staff. We extend special thanks to those of Hawaiian ancestry on whose sacred mountain of Maunakea we are privileged to be guests.}

\bibliography{rgbplanocc}

\end{document}